\documentclass  [12pt] {article}

\begin{document}

\title{%
Phase spaces of Doubly Special Relativity }

\author{A. B\l{}aut\thanks{email: ablaut@ift.uni.wroc.pl},
M. Daszkiewicz\thanks{email: marcin@ift.uni.wroc.pl},\\
J. {Kowalski--Glikman}\thanks{email: jurekk@ift.uni.wroc.pl; Research  partially supported
by the    KBN grant 5PO3B05620.}~, and S. Nowak\thanks{email:
pantera@ift.uni.wroc.pl}\\  \\ {\em Institute for Theoretical
Physics}\\ {\em University of Wroc\l{}aw}\\ {\em Pl.\ Maxa Borna 9}\\
{\em Pl--50-204 Wroc\l{}aw, Poland}} \maketitle

                                                                                                                                                                                                                                                                    \begin{abstract}
We show that depending on the direction of deformation of
$\kappa$-Poincar\'e algebra (time-like,  space-like, or
light-like) the associated phase spaces of single particle in
Doubly Special Relativity theories have the energy-momentum spaces
of the form of de Sitter, anti-de Sitter, and flat space,
respectively.
\end{abstract}
\clearpage

It has been claimed for some time that quantum groups could play
significant role in quantizing gravity (see, for example,
\cite{Major:1995yz} and also  \cite{Bais:2002ye}, where the role
of quantum symmetries in 2+1 quantum gravity has been explicitly
shown.) If this claim is correct in the case of 3+1 quantum
gravity, then one can expect that the symmetries of ``quantum
special relativity'' defined as a flat space limit of quantum
gravity should also be properly defined in terms of quantum
groups.

Following this intuition quantum $\kappa$-Poincar\'e algebra
\cite{kappaP}, \cite{kappaM}  has been proposed some time ago  as
a posible algebra of symmetries of flat space in Planckian regime.
This algebra was incorporated later to the construction of Doubly
Special Relativity (DSR) theories \cite{Amelino-Camelia:2000ge},
\cite{Amelino-Camelia:2000mn}, \cite{Kowalski-Glikman:2001gp},
\cite{Bruno:2001mw}, in which it plays a role of the symmetry
algebra of one-particle states (another version of DSR theory has
been proposed in \cite{Magueijo:2001cr}, \cite{Magueijo:2002am},
but it can be also described in terms of a $\kappa$-Poincar\'e
algebra \cite{Kowalski-Glikman:2002we}.)

It happened to be more natural to introduce DSR as a theory  of
energy and momenta with deformed action of boosts. It turns out,
however that one can make use of the co-product of
$\kappa$-Poincar\'e algebra to construct position sector of DSR.
In particular it has been shown in \cite{Kowalski-Glikman:2002jr},
\cite{Kowalski-Glikman:2002ft}, \cite{Kowalski-Glikman:2003we}
that in DSR one particle dynamics takes place in the phase space,
whose energy-momentum manifold forms four dimensional de Sitter
space, while the space of positions is the non-commutative
$\kappa$-Minkowski space \cite{kappaM}. The link between $\kappa$-Minkowski space and $\kappa$-Poincar\'e algebra, which was originally obtained by using duality of Hopf algebra  (\cite{kappaM}, \cite{zakrz}), has been recently re-derived in the paper
\cite{Agostini:2003vg} 
starting from dynamics on $\kappa$-Minkowski space. 

In the analysis reported  in these papers the starting  point was
a particular form of $\kappa$-Poincar\'e algebra, with deformation
along time-like direction. It is well known however that there are
other $\kappa$-Poincar\'e algebras \cite{kosinski1995},
\cite{Kosinski:2003xx} in which the deformation can be directed
along light-like or space-like directions. In this paper we will
present construction of phase spaces in such a general case.
\newline

Let us start with reviewing the general form of
$\kappa$-Poincar\'e algebra  (in the case of arbitrary Minkowski
metric $g^{\mu\nu}$), whose algebraic part is given by
\cite{kosinski1995}

\begin{eqnarray}
\label{1}
 [M^{\mu \nu},M^{\alpha \beta}] &=& i\left (g^{\mu \beta}M^{\nu
\alpha} - g^{\nu \beta}M^{\mu \alpha} + g^{\nu \alpha}M^{\mu
\beta} - g^{\mu \alpha}M^{\nu \beta} \right )
\\
\label{2}
 [P_{\mu},P_{\nu}] &=& 0, \quad [M^{ij},P_{0}] = 0
\\
\label{3} [M^{ij},P_{k}] &=&
i\kappa\left(\delta^{j}_{k}g^{0i}-\delta^{i}_{k}g^{0j}\right)\left(1-{\rm
e}^{-P_0/\kappa}\right) +
i\left(\delta^{j}_{k}g^{is}-\delta^{i}_{k}g^{js}\right)P_{s}
\\
\label{4}
  [M^{i0},P_{0}] &=& i\kappa g^{i0}\left(1-{\rm e}^{-P_0/\kappa}\right) +
ig^{ik}P_{k}
\\
\nonumber [M^{i0},P_{k}] &=&
-i\frac{\kappa}{2}g^{00}\delta^{i}_{k}\left(1-{\rm
e}^{-2P_0/\kappa}\right) - i\delta^{i}_{k}g^{0s}P_s{\rm
e}^{-P_0/\kappa}
\\
\label{5} && + ig^{0i}P_k\left({\rm e}^{-P_0/\kappa} - 1\right) +
\frac{i}{2\kappa}\delta^{i}_{k}g^{rs}P_rP_s -
\frac{i}{\kappa}g^{is}P_sP_k
\end{eqnarray}

\noindent Here Greek indices run from $0$ to $D$, where $D+1$ is
the dimension of  spacetime under consideration, while Latin from
1 to $D$. It is easy to see that the standard $\kappa$-Poincar\'e
algebra of DSR in the bicrossproduct basis is a particular example
of the general algebra (\ref{1})--(\ref{5}), corresponding to the
metric $g_{\mu\nu} = \mbox{diag}(-1,1,1,1)$ (in four spacetime
dimensions) so that the deformation is in the time-like direction.
One can consider other cases, however, with $g_{\mu\nu} =
\mbox{diag}(1,-1,1,1)$, and
\begin{equation}\label{5a}
 g_{\mu\nu}=\left(
\begin{array}{cccc}
  0 & 1 & 0 & 0 \\
  1 & 0 & 0 & 0 \\
  0 & 0 & 1 & 0 \\
  0 & 0 & 0 & 1 \\
\end{array}\right)
\end{equation}
corresponding to space-like and light-like \cite{kosinski1995},
\cite{Kosinski:2003xx} deformation, respectively.  It should be
stressed that analogous construction can be made in any dimension
$D\geq1$.

According to \cite{kosinski1995} the algebra $(1)$--$(5)$ with
the following co-product structure
\begin{eqnarray}
\label{6}
 \Delta (P_0) &=& 1\otimes P_0 + P_0 \otimes 1
\\
\label{7}
\Delta (P_k) &=&
P_k\otimes {\rm e}^{-P_0/\kappa} + 1 \otimes P_k
\\
\label{8}
  \Delta
(M^{ij}) &=& M^{ij}\otimes 1 + 1 \otimes M^{ij}
\\
\label{9}
  \Delta
(M^{i0}) &=& 1\otimes M^{i0} + M^{i0} \otimes {\rm
e}^{-P_0/\kappa} - \frac{1}{\kappa}M^{ij}\otimes P_j
\end{eqnarray}
and appropriately given antipode defines Hopf algebra, which we
call quantum $\kappa$-Poincar\'e algebra with arbitrary metric
$g_{\mu\nu}$. To extend this structure to the whole of the phase
space of a one-particle system one adds the dual quantum group
with generators being Lorentz transformations $\Lambda^\mu{}_\nu$
and translations of momenta, which can be interpreted as positions
$X^\mu$.  Their co-products are \cite{kosinski1995}
$$
 \Delta(X^{\mu})=\Lambda^{\mu}{}_{\nu}\otimes
X^{\nu}+X^{\mu}\otimes 1
$$
and
$$
\Delta(\Lambda^{\mu}{}_{\nu})=\Lambda^{\mu}{}_{\rho}\otimes
\Lambda^{\rho}{}_{\nu}
$$
Next one defines the pairings between elements of the algebra  and
the group as follows
\begin{eqnarray}
\nonumber
 \left<P_{\mu},X^{\nu}\right>&=& i\delta_{\mu}^{\nu}
\\
\nonumber
  \left<\Lambda^{\mu}{}_{\nu},M^{\alpha\beta}\right>&=& i
\left(g^{\alpha\mu}\delta^{\beta}_{\nu}-g^{\beta\mu}\delta^{\alpha}_{\nu}\right)
\\
\nonumber
  \left<\Lambda^{\mu}{}_{\nu},1\right>&=& \delta^{\mu}_{\nu}
\end{eqnarray}

The phase space brackets can be found by employing   Heisenberg
double procedure \cite{crossalg}, \cite{luno},
\cite{Kowalski-Glikman:2002jr}
\begin{eqnarray}
\nonumber
 \left[X^{\mu},P_{\nu}\right]&=& P_{\nu(1)}\left<X^{\mu}_{(1)},P_{\nu(2)}\right>X^{\mu}_{(2)}
-P_{\nu}X^{\mu},
\\
\nonumber
  \left[X^\mu,M^\rho{}_\sigma\right]&=& M_{(1)}^\rho{}_\sigma \left< X_{(1)}^\mu,M_{(2)}^\rho{}_\sigma \right> X_{(2)}^\mu - M^\rho{}_\sigma X^\mu,
\end{eqnarray}
where we make use of the standard (``Sweedler'') notation for co-product
$$
\Delta {\cal
T} = \sum {\cal T}_{(1)} \otimes {\cal T}_{(2)}.
$$
In this way one gets the following phase space commutator
constituting the so called $\kappa$-Minkowski,  non-commutative
spacetime
\begin{eqnarray}
\label{17}
[X^{\mu},X^{\nu}]&=&-\frac{i}{\kappa}X^{\mu}\delta^{\nu}_{0}+
\frac{i}{\kappa}X^{\nu}\delta^{\mu}_{0}
\\
\label{18}
[X^{\mu},M^{k0}]&=&i\left(g^{k\mu}X^{0}-g^{0\mu}X^{k}\right)
-\frac{i}{\kappa}\left(\delta_{0}^{\mu}M^{k0}+\delta_{l}^{\mu}M^{kl}\right)
\\
\label{19}
[X^{\mu},M^{kl}]&=&i\left(g^{k\mu}X^{l}-g^{l\mu}X^{k}\right)
\end{eqnarray}
and the commutators between positions $X$ and momenta
\begin{equation}
\label{px}
\left[P_{\nu},X^{\mu}\right]=-i\delta^{\mu}_{\nu}+\frac{i}{\kappa}\delta_{0}^{\mu}P_{\sigma}\left(
\delta^{\sigma}_{\nu}-\delta^{0}_{\nu}\delta^{\sigma}_{\nu}\right).
\end{equation}

It is interesting to ask which algebra the generators $X^\mu$ and
$M^{\mu\nu}$ form.  Because of (\ref{2})--(\ref{5}) and (\ref{px}) this $(D+1)(D+2)/2$-dimensional algebra is the
algebra of symmetries of the $D+1$ dimensional energy-momentum
space. Assuming that the symmetries act on the energy-momentum
space transitively, and that the subalgebra of symmetries leaving
invariant the point $P_\mu=0$ is the Lorentz subalgebra, we see
that energy-momentum space is isomorphic to the quotient of the
group generated by $X$ and $M$ algebra by its Lorentz subgroup.
Therefore knowing the algebra of $X$ and $M$ we know the form of
energy-momentum manifold.

To find the algebra in question let us make the identification
$$
 {\hat M}^{\mu \nu} = M^{\mu \nu}, \quad {\hat M}^{(D+1) \mu}=-{\hat M}^{\mu(D+1)} ={\kappa}
X^{\mu}
$$
One can easily check that the algebra of $(D+1)(D+2)/2$ generators ${\hat M}^{AB}$, ($A,B = 0, \ldots D+1 $) satisfying
\begin{equation}\label{20}
  [{\hat M}^{AB},{\hat M}^{CD}] = i\left ({\hat g}^{BC}{\hat
M}^{AD} - {\hat g}^{AC}{\hat M}^{BD} - {\hat g}^{BD}{\hat M}^{AC}
+ {\hat g}^{AD}{\hat M}^{BC} \right )
\end{equation}
is equal to  (\ref{17})--(\ref{19}) if
$$
  {\hat g}^{AB}=\left(
\begin{array}{ccccc}
  g^{\mu\nu}         &
\begin{array}{ccc}
  1\\
  0\\
  \vdots\\
  0\\
  \end{array}
\\
  1\;\;\;\;0\cdots 0   &      0         \\
  \end{array}
\right)
$$
where
$$
  g^{\mu\nu}=
\left(
\begin{array}{cccc}
  g^{00} & \cdots & g^{0D} \\
  \vdots & \ddots & \vdots \\
  g^{D0} & \cdots & g^{DD} \\
\end{array}
\right)
$$
is the Minkowski spacetime metric  in the Hopf algebra
(\ref{1})--(\ref{9}).

First we ask if the metric ${\hat g}^{AB}$ can be degenerate,
i.e.,  if it can have zero eigenvalues. Since the metric
$g^{\mu\nu}$ is non-degenerate (with eigenvalues $-1, 1, \ldots,
1$) from standard theorems of matrix algebra it follows that the
rank of ${\hat g}^{AB}$ can be at least $D+1$, and thus it can
have at most one zero eigenvalue. This happens if, for example, one
of the first $D+1$ columns of ${\hat g}^{AB}$ is proportional to
its last column, i.e., in the light-like case. Straightforward algebraic considerations 
show next that
$\hat g^{AB}$ can have only the following signatures:
$[-,+,\ldots, +]$, $[-,-,+, \ldots, +]$ or $[0,-,+,\ldots, +]$ corresponding to the 
time-like, space-like and light-like deformations, respectively.
\newline

Let us now make this explicit in the four-dimensional case.
\newline

{\bf Time-like deformation}. In this case $g^{\mu\nu} = \mbox{diag
}(-1,1,1,1)$  and the algebra (\ref{20}) is the $SO(1,4)$ algebra,
while the energy-momentum space is de Sitter space
$SO(1,4)/SO(1,3)$. This case has been analyzed before in
\cite{Kowalski-Glikman:2002ft} and \cite{Kowalski-Glikman:2003we}.
\newline

{\bf Space-like deformation}. $g^{\mu\nu} = \mbox{diag
}(1,-1,1,1)$  and the algebra (\ref{20}) is the $SO(2,3)$ algebra.
The energy-momentum space is anti-de Sitter space
$SO(2,3)/SO(1,3)$. It is worth noticing that anti-de Sitter space
is the energy-momentum manifold of 2+1 dimensional quantum gravity
coupled to point particle \cite{Matschull:1997du}. Explicitly, the
relation between the generators $J^\mu$ (rotation and boosts) and
$y^\mu$ (translations) employed in that paper, which satisfy the
algebra
\begin{eqnarray*}
[y^{\mu},y^{\nu}] &=& 2\epsilon^{\mu\nu}{}_{\sigma}y^{\sigma} \cr
[J^{\mu},J^{\nu}] &=& -\epsilon^{\mu\nu}{}_{\sigma}J^{\sigma} \cr
[J^{\mu},y^{\nu}] &=& -\epsilon^{\mu\nu}{}_{\sigma}y^{\sigma}
\end{eqnarray*}
and our generators is the following
\begin{eqnarray*}
J^1 &=& -i\kappa X^0 \cr
J^2 &=& iM^{01}     \cr
J^0 &=& i\kappa X^1 - iM^{01} \cr
y^1 &=& i\kappa X^0 + iM^{12}\cr
y^2 &=& -i\kappa X^2 - iM^{01} +iM^{02}\cr
y^0 &=& -i\kappa X^1 + iM^{01} -M^{02}
\end{eqnarray*}
The relation of 2+1 quantum gravity to DSR theories and its
possible  relevance for 3+1 dimensional physics has been analyzed
in \cite{Amelino-Camelia:2003xp} and \cite{Freidel:2003sp}.
\newline

{\bf Light-like deformation}. Now the metric $g^{\mu\nu}$ is given
by (\ref{5a}) and  it is clear that the metric $\hat g^{AB}$ is
degenerate. In context of DSR theory such deformation has been
analyzed in \cite{Blaut:2003cx}. It is easy to see that in this
case the algebra (\ref{20}) is the standard Poincar\'e algebra and
that the energy-momentum space is the flat Minkowski space.
Explicitly, the
 four commuting elements ${\cal X}^\mu$
 have the form
$$
{\cal X}^0 = X^0 - \frac1\kappa\, M^{10},\quad {\cal X}^1 = X^1,
\quad {\cal X}^\alpha = X^\alpha - \frac1\kappa\, M^{1\alpha}, \;\; \alpha = 2,3.
$$
This case certainly deserves further studies, as it may correspond
to DSR theories with commuting spacetime. There is a hope
therefore that, for example, the construction of $\kappa$-deformed
field theory \cite{Kosinski:2003xx} might be much simpler here
than in other cases. We will present the results of this
investigations in a separate paper.
\newline

In conclusions let us recapitulate the results of reported above.
We  found that in general case, there are three DSR phase spaces
associated with $\kappa$-Poincar\'e algebra. Let us stress that
since this algebra has the same form in any dimension, our results
hold for any space dimension $D\geq 1$ as well. It is of course an
open problem which of this cases (if any) is a correct setting for
3+1 quantum special relativity. Work in this direction is in
progress.
\newline

{\bf Acknowledgement}. We would like to thank P. \L{}ugiewicz and J.\ Lukierski for discussion and useful comments.

\end{document}